\documentclass[aps, prl, twocolumn, superscriptaddress]{revtex4-1}

\usepackage{amssymb, amsmath, amsthm}
\usepackage{xcolor}
\usepackage{graphicx}
\usepackage[protrusion=true,expansion=true]{microtype}
\usepackage{times}
\usepackage{hyperref}
\hypersetup{
    pdftitle={main},    
    pdfauthor={},
    colorlinks=true,
    citecolor=blue,
    linkcolor=blue,
    urlcolor=blue
  }
\usepackage{comment}
\usepackage{blkarray}
\usepackage{mathtools}
\usepackage{tikz}
\usetikzlibrary{decorations.pathreplacing}
\usepackage{dsfont}

\usepackage{graphicx}
\usepackage{dcolumn}
\usepackage{bm}

\usepackage[makeroom]{cancel}
\usepackage{latexsym}
\usepackage{longtable}
\usepackage{epsfig}
\usepackage{graphicx,bbm,psfrag}
\usepackage{amssymb,amsmath,amsthm,booktabs,mathtools}
\usepackage{bm}
\usepackage{color}
\usepackage{dutchcal}
\usepackage{hyperref}

\hypersetup{
    colorlinks,
    citecolor=red,
    linkcolor=blue,
}

\newcommand{\bc}{\begin{center}}
\newcommand{\ec}{\end{center}}
\def\ba#1{\begin{array}{#1}\displaystyle}
\newcommand{\ea}{\end{array}}

\newcommand{\beq}{\begin{equation}}
\newcommand{\eeq}{\end{equation}}
\newcommand{\beqa}{\begin{eqnarray}}
\newcommand{\eeqa}{\end{eqnarray}}
\newcommand{\no}{\nonumber}
\newcommand{\n}{\nonumber\\}
\newcommand{\bi}{\begin{itemize}}
\newcommand{\ei}{\end{itemize}}

\def\frc#1#2{\frac{#1}{#2}}

\newcommand{\p}{\partial}

\newcommand{\bra}{\langle}
\newcommand{\ket}{\rangle}

\newcommand{\prlsection}[1]{{\em {#1}.---~}}

\newcommand{\ep}{\epsilon}

\newcommand{\dd}{{\rm d}}

\DeclareMathOperator{\sgn}{sgn}

\newcommand{\dbra}{\langle\hspace{-0.17em}\langle}
\newcommand{\dket}{\rangle\hspace{-0.17em}\rangle}

\newcommand{\upd}[1]{^\mathrm{#1}}
\newcommand{\ind}[1]{_\mathrm{#1}}

\newcommand{\titleinfo}{
Diffusive hydrodynamics from long-range correlations
}
\begin{document}

\preprint{APS/123-QED}

\title{\titleinfo
}

\author{Friedrich Hübner}
\affiliation{Department of Mathematics, King’s College London, Strand WC2R 2LS, London, U.K.}

\author{Leonardo Biagetti}
 \affiliation{Laboratoire de Physique Th\'eorique et Mod\'elisation, CNRS UMR 8089,
	CY Cergy Paris Universit\'e, 95302 Cergy-Pontoise Cedex, France}

\author{Jacopo De Nardis}
\affiliation{Laboratoire de Physique Th\'eorique et Mod\'elisation, CNRS UMR 8089,
	CY Cergy Paris Universit\'e, 95302 Cergy-Pontoise Cedex, France}

\author{Benjamin Doyon}
\affiliation{Department of Mathematics, King’s College London, Strand WC2R 2LS, London, U.K.}

\date{\today}

\begin{abstract}
In the hydrodynamic theory, the non-equilibrium dynamics of a many-body system is approximated, at large scales of space and time, by irreversible relaxation to local entropy maximisation. This results in a convective equation corrected by viscous or diffusive terms in a gradient expansion, such as the Navier-Stokes equations. Diffusive terms are evaluated using the Kubo formula, and possibly arising from an emergent noise due to discarded microscopic degrees of freedom. In one dimension of space, diffusive scaling is often broken as noise leads to super-diffusion. But  in linearly degenerate hydrodynamics, such as that of integrable models, diffusive behaviors are observed, and it has long be thought that the standard diffusive picture remains valid. In this letter, we show that in such systems, the Navier-Stokes equation breaks down beyond linear response.
We demonstrate that diffusive-order corrections do not take the form of a gradient expansion. Instead, they are completely determined by ballistic transport of initial-state fluctuations, and obtained from the non-local two-point correlations recently predicted by the ballistic macroscopic fluctuation theory (BMFT); the resulting hydrodynamic equations are reversible. To do so, we establish a regularised fluctuation theory, putting on a firm basis the recent idea that ballistic transport of initial-state fluctuations determines fluctuations and correlations beyond the Euler scale. This extends the idea of ``diffusion from convection'' previously developed to explain the Kubo formula in integrable systems, to generic non-equilibrium settings. 
\end{abstract}

\maketitle
\prlsection{Introduction and main results } Hydrodynamics is one of the most successful theories for describing the dynamics of non-equilibrium complex systems~\cite{Pines1966, PhysRevD.53.5799, gaspard1997, bressan2000hyperbolic, Boldrighini1997, landau_fluid_mechanics, 1805.09331, Lucas2018, PhysRevD.85.085029, Jlicher2018, Le2023, Narozhny2022, Malvania2021, PhysRevE.108.054101, Grozdanov2019}. In recent decades, it has been applied to various physical situations, including quantum many-body theory and strongly correlated particles \cite{castro2016emergent, bertini2016transport, PhysRevD.85.085029, ruggiero2020quantum, Lucas2018, Malvania2021, schemmer2019generalized, Scheie2021, schemmer2019generalized, moller2021, cataldini2022, PhysRevA.107.L061302, bulchandani2018bethe, Bulchandani2020, piroli2017transport}. The hydrodynamic formulation relies on several assumptions, primarily the fluid-cell approximation~\cite{landau_fluid_mechanics, de2023hydrodynamic}, which asserts that the system locally relaxes towards a stationary state. Given the dynamics' integrals of motion \(Q_i = \ell \int \dd x\, q_i(x)\) (here and below we use macroscopic coordinates $x,t = x_{\rm micro}/\ell,t_{\rm micro}/\ell$ where $\ell$ is a large length scale) and their local densities, which satisfy continuity equations \(\partial_t q_i + \partial_x j_i = 0\), the hydrodynamic approximation assumes the existence of fields \(\beta^i(x,t)\) such that the state around each space-time point \(x,t\) is described by an independent, locally relaxed ``fluid cell'' determined by the generalized temperatures $\beta^i(x,t)$. A probability distribution, or density matrix, representing this factorization in space on time-slice \(t\) is (summation over repeated indices implied):
\begin{equation}\label{eq:hydro01}
    \rho \sim \exp \Bigg[-\ell\int \dd x\, \beta^i(x,t) q_i(x) \Bigg],
\end{equation}
which typically has exponentially decaying correlations. Averaging the continuity equation over \eqref{eq:hydro01} one obtains \(\partial_t \langle q_i \rangle_{x,t} + \partial_x \langle j_i \rangle_{x,t} = 0\); here the state \(\langle \cdot \rangle_{x,t}\) is \eqref{eq:hydro01} but with \(\beta^i\) constant, evaluated at \(x,t\). By inverting the relation between $\{\beta_i\}$ and $\{\langle q_i \rangle\}$, one can express the currents $\langle j_i \rangle_{x,t} = \mathcal{j}_i(\langle q_\cdot \rangle_{x,t})$ as function of local density. Denoting $\mathcal{q}_i = \langle q_i(x,t)\rangle$ (which at this order of approximation equals $\langle q_i \rangle_{x,t}$) the so called Euler hydrodynamic equations become
\begin{equation}\label{eq:euler}
    \partial_t \mathcal{q}_i + A_{i}^{~k} \partial_x \mathcal{q}_k =0, 
\end{equation}
where the flux jacobian \(A_i^{~k} = \delta \mathcal{j}_i / \delta \mathcal{q}_k\) describes ballistic transport. They are valid in any many-body systems if the observation length and time scales are sufficiently large. 

Since the Euler equations conserve total entropy, in order to understand phenomena such as thermalization it is necessary to consider corrections beyond them. The hydrodynamic expansion accounts for corrections in terms of the smoothness of the generalized temperatures \(\beta^i(x) \simeq \beta^i(x_0) + (x - x_0) \partial \beta^i + \ldots\) around each point \(x_0\), meaning that in general $\mathcal{q}_i \neq \langle q_i \rangle_{x,t}$. The distribution \eqref{eq:hydro01} does not directly represent these corrections; instead, they follow from the assumption that fast relaxation occurs, on microscopic time scales, from such a local equilibrium state. This yields the well-known Navier-Stokes (NS) equations~\cite{landau_fluid_mechanics, Boldrighini1997, SpohnBook, 10.21468/SciPostPhys.6.4.049, Bulchandani2024}:
\begin{equation}\label{eq:NS}
    \partial_t \mathcal{q}_i + A_{i}^{~k} \partial_x \mathcal{q}_k = \frac{1}{2\ell} \partial_x (\mathfrak{D}_i^{~k} \partial_x \mathcal{q}_k).
\end{equation}
The direct (DC) conductivity (Onsager matrix) \(\mathfrak L_{ij} = \mathfrak{D}_i^{~k} C_{k,j}\), where the susceptibility matrix is \(C_{ik} = \langle Q_i q_k(0)\rangle^c_{x,t}\), is obtained via the Kubo formula \(\mathfrak{L}_{i,j} = \int_{-\infty}^{\infty} \dd t' \int \dd x' \langle j^-_i(x',t') j_j(0,0) \rangle^c_{x,t}\), where the asymptotic ballistic contribution is subtracted out \(j_i^- = j_i - A_i^{~k} q_k\), as stated by the celebrated Einstein relation~\cite{kubo1966}. Kubo diffusion -- Eq.~\eqref{eq:NS} along with the Kubo formula to evaluate $\mathfrak D_{i}^{~k}$ -- is well-justified in linear response theory around an equilibrium state, however, in writing \eqref{eq:NS} one makes the assumption that it remains valid beyond linear response. We will show that this assumption is incorrect, although it is a good approximation at early times from local-equilibrium states \eqref{eq:hydro01}. Throughout this paper we will assume that all densities and currents are Parity-Time  (PT) symmetric; this implies positivity of the Onsager matrix and thus entropy production under the NS equation \eqref{eq:NS} \cite{de2019diffusion,de2023hydrodynamic}.

A description as above holds in arbitrary dimensions, but we focus on one dimension of space. In this case, there is an important caveat. When \textit{at least one interacting ballistic mode is present}, such as in anharmonic chains or interacting one-dimensional gases, Eq. \eqref{eq:hydro01} is typically broken: the matrix $\mathfrak L$ has divergent components, and superdiffusion arises, as demonstrated in the framework of non-linear fluctuating hydrodynamics~\cite{Spohn2014, Spohn2015-2, Doyon2022, Spohn2016, Mendl2016, Popkov2015, PhysRevLett.120.240601}. In certain systems, such as integrable models, it is finite~\cite{spohn_large_scale_dynamics, 10.21468/SciPostPhys.6.4.049, PhysRevB.98.220303, PhysRevLett.121.160603, Doyon2022, medenjak2020diffusion} (with notable exceptions in isotropic spin chains~\cite{de2020superdiffusion, PhysRevLett.122.210602, Wei2022, Scheie2021, PhysRevE.100.042116, PhysRevLett.131.197102}). As we will show in the SM (see remark after Eq. (S35)),  finiteness of $\mathfrak L$ in one-dimensional systems follows if the hydrodynamic equation has a property called \emph{linear degeneracy} \cite{FERAPONTOV1991112,lindeg1,lindeg2}; generalised hydrodynamics (GHD), for integrable systems, has this property \cite{el2011kinetic, pavlov2012generalized, PhysRevB.101.041411, Doyon2020}. This property implies that hydrodynamic modes do not ``self-interact"; only interactions between modes with different velocities are present. Self-interaction is the source of super-diffusion, as non-linear hydrodynamics is unstable under small noise in one dimension~\cite{PhysRevLett.56.889}. Thus, the Kubo-diffusion understanding is that in linearly degenerate systems, Eq. \eqref{eq:NS} is expected to describe the space-time profiles of all one-point averages of local observables \(\langle o(x,t) \rangle\); and we will explain how this must be corrected.

Although the approximation Eq.~\eqref{eq:hydro01} accurately describes one-point averages at the Euler scales, recent findings via the ballistic macroscopic fluctuation theory (BMFT)~\cite{PhysRevLett.131.027101, 10.21468/SciPostPhys.15.4.136} indicate that different fluid cells do not remain exponentially uncorrelated at all times during non-equilibrium dynamics.
It is found~\cite{PhysRevLett.131.027101} that correlations of order $1/\ell$ appear on macroscopic distances at any positive macroscopic time if the model admits hydrodynamic modes with at least two different velocities and is interacting. In the two-point functions \(S_{ij}(x,y,t) = \ell \langle q_i(x,t) q_j(y,t) \rangle^c \), in addition to the $\delta$-contribution due to microscopic correlations described by \eqref{eq:hydro01}, there is a regular contribution $E_{ik}(x,y,t)$ representing these long-range correlations:
\begin{equation}\label{eq:shapeofCorr}
    S_{ij}(x,y,t) = \delta(x-y) C_{ij}(x,t) + E_{ij}(x,y,t).
\end{equation}
As the Einstein relation connects fluctuations (and correlations) to diffusion, it is reasonable to ask about the relation between these newly found long-range correlations and the diffusive order of hydrodynamics.

In this letter, we propose a \emph{new theory for the diffusive order of hydrodynamics}, expected to hold in all one-dimensional systems with linear degeneracy, such as integrable models. It is not based on Kubo-type relaxation but on the principle that initial fluctuations propagate deterministically and ballistically, according to the Euler equation. For Euler-scale fluctuations, this principle has its roots in~\cite{DMfluctuballistic2020} and is encoded within the ballistic macroscopic fluctuation theory (BMFT) \cite{10.21468/SciPostPhys.15.4.136}. Here, we assume this principle holds \emph{up to, including, diffusive scales}, for instance for corrections of order $1/\ell$ in local expectation values. This has recently been proposed to describe anomalous fluctuations \cite{PhysRevB.109.024417, PhysRevLett.128.160601, yoshimura_anomalous_2024, non_gaussian_diffusive_dirac_fluids}. According to this principle, we show that the \textit{NS equation \eqref{eq:NS} is not the correct equation for large-scale dynamics} up to diffusive order. We derive the new equation, where the diffusive-order term is replaced by a term determined by the BMFT long-range correlations $E_{ik}(x,y,t)$. For linear perturbations to homogeneous, stationary states, this is still in agreement with the Navier-Stokes equation \eqref{eq:NS}, but it otherwise gives different predictions.

We verify our new equation through Monte Carlo simulations and first-principle calculations~\cite{UPCOMING} in the hard rods gas \cite{Flicker68, Robledo1986, Doyon2017, PhysRevE.108.064130}. 

\prlsection{BMFT and equation for diffusive hydrodynamics}
The BMFT can be formulated as a theory for fluctuating conserved densities \cite{10.21468/SciPostPhys.15.4.136}. One sets a measure on their initial configurations $q_i^0(x)$, which are then evolved deterministically according to the Euler equation: $q_i(x,0) = q_i^0(x)$ and $\p_t q_i(x,t) + \p_x \mathcal j_i(q_\cdot(x,t))=0$. In linearly degenerate systems, and in particular in GHD, the solution is unique (in particular, no shocks appear \cite{FERAPONTOV1991112,lindeg1,lindeg2,hübner2024newquadraturegeneralizedhydrodynamics}) and thus $q_i(x,t)=q_i[q^0](x,t)$ is a fixed functional of $q_\cdot^0(\cdot)$: initial fluctuations are deterministically transported. Further, any observable at $x,t$ is taken to be a fixed function of conserved densities at $x,t$, simply its expectation value in the corresponding maximal entropy state, $a(x,t) \equiv \mathcal a(q_\cdot(x,t))$. That is, for one-point averages, $\bra a(x,t)\ket \sim \dbra \mathcal a(q_{\cdot}(x,t))\dket$ with 
\beq\label{eq:BMFT}
	 \dbra \mathcal a(q_{\cdot}(x,t))\dket := \int [\dd q^0]\,e^{-\ell\mathcal F_{\rm ini}[q^0]} \mathcal a(q_{\cdot}[q^0](x,t)),
\eeq
where $-\mathcal F_{\rm ini}[q]$ is an entropy functional describing the initial state. In a local GGE \eqref{eq:hydro01}, one has $-\mathcal F_{\rm ini}[q^0] = \int \dd x\,s[q_\cdot^0(x)|\beta^\cdot(x)]$, for a relative entropy density $s$ of the fluctuating state regarding the target local GGE (see \cite{10.21468/SciPostPhys.15.4.136}), which gives $S_{ik}(x,y,0) = \delta(x-y) C_{ik}(x,0)$, the delta-function part of \eqref{eq:shapeofCorr}. This delta-function part in $S_{ik}(x,y,0)$ represents the finite correlations of conserved densities $q_i(x,t)$'s on microscopic lengths, whose exact shapes are lost in this approximate theory and that are re-scaled to infinity as $\ell\to\infty$ because of the factor $\ell$ in the definition of $S_{ik}(x,y,t)$. In BMFT, Eq.~\eqref{eq:BMFT} holds at leading order in $\ell^{-1}$, thus one takes the $\ell\to\infty$ limit which concentrates the integral on its saddle point $\mathcal q_i(x,t)=\bra q_i(x,t)\ket$. This gives $\bra a(x,t)\ket = \mathcal a(\mathcal q_\cdot(x,t))$ and the Euler hydrodynamic equation $\p_t \mathcal q_i(x,t) + \p_x \mathcal j(\mathcal q_\cdot(x,t))=0$. Similarly, one obtains the evolution equation for the two-point functions \cite{10.21468/SciPostPhys.5.5.054,10.21468/SciPostPhys.15.4.136}
\begin{equation}\label{eq:timeevCorr0} 
\partial_t S_{ij} + \partial_x \big(A_i^{~k} S_{kj}\big) + \partial_y \big(S_{ik} A_j^{~k}\big) = 0.
\end{equation}

{\em We now claim that Eq.~\eqref{eq:BMFT} correctly describes the microscopic evolution up to, including, order $1/\ell$}: namely, by keeping the $1/\ell$ fluctuations around the saddle point, \eqref{eq:BMFT} also gives the GHD diffusive terms beyond Euler evolution. In this case, one-point averages $\bra a(x,t)\ket$ (and higher-point functions such as \eqref{eq:shapeofCorr}) receive corrections of order $1/\ell$. In the following, we shall still denote $\mathcal q_i(x,t) = \bra q_i(x,t)\ket$, including all corrections, with evolution equation $\p_t \mathcal q_i(x,t) + \p_x \bra j_i(x,t)\ket = 0$.  We seek the $1/\ell$ corrections to $\bra j_i(x,t)\ket$ (or more generally $\bra a(x,t)\ket$) which will give rise to the hydrodynamic equation up to diffusive order.

Since BMFT is formally not a theory on microscopic observables, but a theory on fluid-cell averaged quantities, we need to regularize observables by doing an appropriate fluid cell averaging, denoted by $[a]^{\rm{reg}}$, see Eq.~\eqref{eq:regularisation} in the End Matter. For simplicity at $t=0$ we assume an uncorrelated local GGE state \eqref{eq:hydro01}: in this state we have $\dbra \mathcal [a(q_\cdot(x,0))]^{\rm reg}\dket = \mathcal a(\mathcal q_\cdot(x,0))$ and if the observable is PT-symmetric $\mathcal a(\mathcal q_\cdot(x,0)) = \bra a(x,0)\ket + O(\ell^{-2})$. We concentrate on such observables.

The next step is to expand \eqref{eq:proposal} via a {\em cumulant expansion}, which takes the general form (for expectation $\bra\cdots\ket$ over some random variable $q$)
\begin{equation}
  \bra f(q)\ket = f(\bra q\ket) + (1/2)f''(\bra q\ket)\bra q^2\ket^{\rm c}+ \ldots  .
\end{equation}
Applied to $\dbra\cdots\dket$, this gives
\begin{equation}\label{eq:cumulantonepoint}
\dbra [\mathcal a(q_\cdot(x,t))]^{\rm reg}\dket = \mathcal a + \frac{1}{2} \frac{\delta \mathcal{a}}{\delta \mathcal{q}_i \delta \mathcal{q}_j}\,\ell^{-1}E^{\rm sym}_{ij}(x,t)+O(\ell^{-2}),
\end{equation}
where the regularisation $[\cdots]^{\rm reg}$ gave rise to the symmetric point-splitting of the correlation $E^{\rm sym}_{ij} = ({E_{ij}(x^+, x^-; t) + E_{ij}(x^-, x^+; t)})/{2}$. From this, we finally obtain the diffusive-order hydrodynamic equation, 
\begin{equation}\label{eq:mainresult}\boxed{
    \partial_t \mathcal{q}_i + A_i^{~k} \partial_x \mathcal{q}_k + \partial_x \Big( \frac{1}{2} \frac{\delta   \mathcal{j}_i(\mathcal q_{\cdot})   }{{\delta \mathcal{q}_r \delta \mathcal{q}_k}} \,\ell^{-1}E^{\rm sym}_{rk} \Big) = 0,}
\end{equation}
which is our main result.

Although Eq.~\eqref{eq:mainresult} gives the diffusive order $\ell^{-1}$ correction to the Euler equation, it is {\em not} a diffusion equation. Clearly, it is also not closed: the $\ell^{-1}$ order involves the symmetric part of the long-range correlations. However, the long-range correlations found from the Euler-scale BMFT, $\ell^{-1}E^{\rm sym}_{rk}$, are already at order $\ell^{-1}$, hence they solve the BMFT equation \eqref{eq:timeevCorr0}. Thus, \eqref{eq:mainresult} and \eqref{eq:timeevCorr0} form a closed system. We may extract from \eqref{eq:timeevCorr0} the equation for the long-range part in \eqref{eq:shapeofCorr} (see \cite{SM}), in which the delta-correlation part acts as a source term:
\begin{align}\label{eq:timeevCorr}
    \p_t (\ell^{-1} E_{ij}) +  & \p_x \big(A_i^{~k} \ell^{-1} E_{kj}\big)
    +\p_y \big ( \ell^{-1} E_{ik}A_j^{~k}\big) \nonumber 
    \\& = - \ell^{-1} \delta(x-y)  \langle j_l^-,  q_i,q_j \rangle  C^{lk} \p_x \mathcal{q}_k  ,
\end{align}
where $\bra a,b,c\ket = \int \dd x'\dd x''\,\bra a(x')b(x'')c(0)\ket^{\rm c}$ is the ``three-point coupling'' and $C^{lk}$ is the inverse of the susceptibility matrix, $C^{lk}C_{ki} = \delta_i^l$ (both evaluated in the GGE represented by $\mathcal q_\cdot(x,t)$). Therefore, the full system of hydrodynamic equations up to (including) the diffusive order is \eqref{eq:mainresult} with \eqref{eq:timeevCorr}.

\noindent Remark 1: \eqref{eq:mainresult} and \eqref{eq:timeevCorr} still contain the length scale $\ell$. In fact, they are invariant under rescaling of $\ell$ while keeping microscopic lengths the same: $\ell^{-1}E_{kj}$ are the physical, long-wavelength, long-range correlations measured just beyond the region of strong microscopic-length correlations, and $\ell^{-1}\delta(x-y)=\delta(x_{\rm micro}-y_{\rm micro})$. With this understanding, one could set $\ell = 1$. However, \eqref{eq:mainresult} and \eqref{eq:timeevCorr} are asymptotic results, valid, in a coarse-grained way, as an asymptotic expansion in large $x_{\rm micro},t_{\rm micro}$ and large initial state wavelengths.

\noindent Remark 2: \eqref{eq:mainresult} seems to be reminiscent of the BBGKY approach, where the equation for the one-point function depends on the two-point function. The important difference is that in BBGKY one needs to insert the microscopic two-point function, while $E^{\rm sym}_{rk}$ are coarse-grained long-range correlations.

It is possible to write the state at any time $t$ in a form similar to \eqref{eq:hydro01}, which fully accounts both for long-range two-point functions \eqref{eq:shapeofCorr} at leading order in $\ell^{-1}$ (BMFT), and one-point function at leading $O(1)$ and first subleading $O(\ell^{-1})$ order (in the regularised representation \eqref{eq:proposal}). Indeed, we show \cite{SM} that the following ``non-local GGE'' implements this:
\begin{align}\label{eq:nonlocalgge}
	\rho_{\rm LR} \propto &\exp \Big[-\int \dd x\, \beta^i(x) q_i(x)+\\  & \ +  \int_{x\neq y} \hspace{-0.3cm}\dd x \dd y\, \frc{\gamma^{ij}(x,y)}2 \delta {q}_i(x)  \delta {q}_j(y) \Big],\nonumber
\end{align}
with the centered fluctuation $\delta {q}_i = q_i - \mathcal{q}_i$. Here, $\beta^\cdot$ fix the values of the charges $\mathcal q_\cdot$, and the matrix $\gamma^{ij}(x,y,t)$ determines the values of the correlations $E^{ij}(x,y,t)$. Note how this natural representation of the long-range correlations, via the $\gamma$-term, directly implements the regularisation \eqref{eq:proposal}, providing conceptual evidence for its validity.

The idea of representing currents in terms of fluctuations of charges constitutes the basis of the theory of nonlinear fluctuating hydrodynamics \cite{Spohn2014,Mendl2016,Spohn2015-2}.
It was also used to evaluate the Onsager matrix in integrable models (and other linearly degenerate models), where the current is projected onto quadratic charges \cite{PhysRevB.98.220303,medenjak2020diffusion,Doyon2022}. The latter approach was dubbed ``diffusion from convection'' \cite{medenjak2020diffusion}, and shown in general to give a lower bound to the diffusion matrix \cite{Doyon2022} and to be the source of anomalous fluctuations in generic systems with at least one ballistic mode \cite{Spohn2015-2,non_gaussian_diffusive_dirac_fluids,mcculloch2024}. 

In this letter, we go further: we propose that extending the BMFT to a fluctuation theory up to the diffusive scale, gives not only the diffusive-scale linear-response equation and anomalous fluctuations, but also, by the cumulant expansion and the regularisation \eqref{eq:regularisation}, the full hydrodynamic equation.

\prlsection{Relation to Navier-Stokes theory}
The (linearized version of the) NS equation \eqref{eq:NS} is well-justified in the linear response regime. In the end matter we derive that \eqref{eq:mainresult} and \eqref{eq:timeevCorr} reduces to \eqref{eq:NS} within linear response theory. Furthermore, we also derive that \eqref{eq:NS} is valid at infinitesimal macroscopic time $t = 0^+$. This must hold as the initial state \eqref{eq:hydro01} is locally close to equilibrium and thus linear response \eqref{eq:NS} describes early times (this was proven in the gas of hard rods~\cite{Boldrighini1997}).

These results show that our new theory reduces to \eqref{eq:NS} close to equilibrium, but supersedes it for the non-equilibrium states which arise throughout the evolution.

\prlsection{Entropy conservation and reversibility} Under PT symmetry, the entropy $-\int \dd x\,{\rm Tr}(\rho(x,t)\log \rho(x,t))$ is strictly non-decreasing in $t$ under the NS evolution \eqref{eq:NS}, as a result of the positivity of the Onsager matrix \cite{Doyon2020,de2023hydrodynamic}. Under the new diffusive-order hydrodynamic equations \eqref{eq:mainresult}, \eqref{eq:timeevCorr}, this is no longer the case beyond early times. However, with long-range correlations, a more natural entropy is $- {\rm Tr}(\rho_{\rm LR}(t) \log \rho_{\rm LR}(t))$ using \eqref{eq:nonlocalgge}. We find that this is constant in time \cite{SM}: there is no information loss as we consider all relevant correlations, at the required scale, generated under the evolution. Eqs.~\eqref{eq:mainresult} and \eqref{eq:timeevCorr} are, in fact, reversible, as they are PT symmetric, in contrast to the NS equation \eqref{eq:NS}. This has important consequences. Indeed, assume that PT symmetry at the Euler scale is that induced by the symmetry of an underlying microscopic dynamics. Then forward microscopic evolution followed by backward evolution at the same time gives the same macrostate, in a way that is {\em robust to errors}, as long as such errors keep not only local averages, but also long-range two-point correlation functions unchanged. \textit{This is in contrast with generic non-integrable systems}, where hydrodynamic shocks can be formed during time evolution; then information inexorably flows to microscopic scales (higher and higher multipoint correlations), and the mere knowledge of the two-point functions is not enough to be able to fully revert time evolution.

\prlsection{Hard rod simulations}
\begin{figure}[t]
\includegraphics[width=\linewidth]{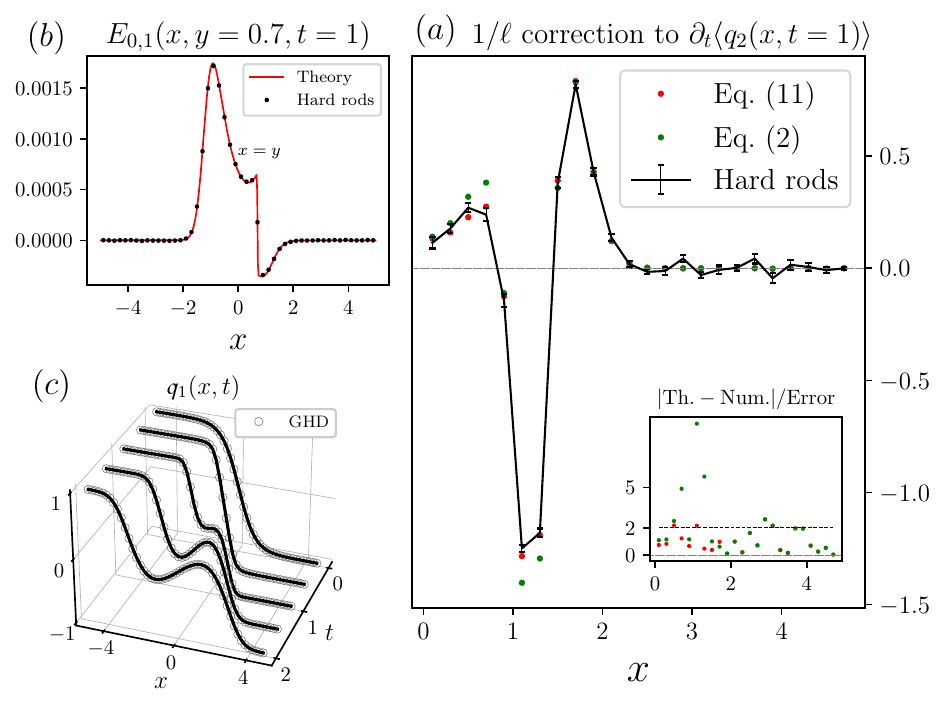}
\caption{Numerical checks for a gas of hard rods with length $a=0.3$: (a) $1/\ell$ correction to $\partial_t\langle q_2(x,t)\rangle$ evaluated at the macroscopic time $t=1$, as a function of the macroscopic position $x$. The initial state is $\rho(x,\theta,t=0)=\exp[-(\theta+\tanh(x))^2/2\sigma]/\sigma\sqrt{2\pi}$. The solid line represents the hard rods data.
More precisely, we measure $\partial_t \mathcal{q}_2(x,t;\ell)\simeq(\mathcal{q}_2(x,t+\Delta t;\ell)-\mathcal{q}_2(x,t-\Delta t;\ell))/2\Delta t$ for $\ell\in\{500,600,...,1000\}$, using $\Delta t=0.05$ and averaging over $8\times 10^{10}$ realizations for each value of $\ell$. Subsequently, at each fixed macroscopic point $x$, we perform a fit of $\partial_t\langle\mathcal{q}_2(x,t;\ell)\rangle$ with model $f(\ell)=f_1+f_2/\ell$. The plot shows the function $f_2(x)$, and the associated error is estimated as the standard deviation of the parameter of the fit.
The red (green) dots represent the theoretical prediction of Eq. 
\eqref{eq:mainresult} (Eq. \eqref{eq:NS}).
The inset on the bottom right shows the distance between the numerics and the theoretical predictions, normalized with the numerical errors.
(b) Regularized correlation $E_{0,1}^{\rm sym}$ at macroscopic time $t=1.0$ between points $x$ and $y=0.7$ in the hard-rod simulations (black dots) compared with the analytical prediction (red line). Note in particular the discontinuity at $x=y$.
(c) Dynamics of the momentum density $\mathcal{q}_1(x,t)$ (solid line), compared with the Euler scale GHD (dots).
}
\label{fig: Diffusive correction}
\end{figure} 
We illustrate the results for the paradigmatic hard-rod model. It consists of a one-dimensional gas of $N$ billiard balls of diameter $a$ and unit mass $m=1$. 
The rods have ordered positions $\{x^{\rm micro}_i\}_{i=1}^N$ and move freely with velocities $\{\theta_i\}_{i=1}^N$. The rods collide elastically whenever $x^{\rm micro}_{i+1}-x^{\rm micro}_i=a$, simply exchanging their velocities. 
This model is integrable, since all the initial velocities are preserved by the dynamics \cite{SpohnBook, Doyon2017}. 
In particular, the charge densities are explicitly defined as $\mathcal{q}_i(x,t)=\int\dd{\theta}\,\theta^i \rho(x,\theta,t)$, where $\rho(x,\theta,t)$ is the density of particles in space and rapidity .

We focus on the dynamics of the hard-rod gas with initial rapidity density distribution
$\rho(x,\theta,t=0)=\exp[-(\theta+\tanh(x))^2/2\sigma]/\sigma\sqrt{2\pi}$, 
where $x=x^{\rm micro}/\ell$ and $t=t^{\rm micro}/\ell$ are macroscopic time and space coordinates. This state represents a smoothened version of the sharp partition protocol, where the average particle velocity changes from $1$ at $x=-\infty$ to $-1$ at $x=\infty$.  Note that the density of particles in the initial state is homogeneous $\mathcal{q}_0(x,0)=1$. In a local GGE state with constant particle density $\mathcal q_0$, the hard rods' positions do not depend on the momenta and are given by a Poisson point process in the volume-excluded coordinates $\hat{x}_i = x_i-ai$ \cite{Boldrighini1997}. Thus, we can generate initial particle positions as $x_{i+1}^{\rm micro}=x_i^{\rm micro}+a+(1/\mathcal{q}_0-a)\xi_i$, where $\xi_i$ are i.i.d. standard exponentially distributed variable. The rapidities are then chosen randomly according the to the local rapidity distribution. The dynamics of the momentum distribution $\mathcal{q}_1(x,t)$ is shown in Fig. \ref{fig: Diffusive correction} c).

In Fig. \ref{fig: Diffusive correction} a) we show the comparison between the $1/\ell$ correction to $\partial_t \langle q_2(x,t)\rangle$ measured from hard rods simulations and the theoretical predictions from formulas \eqref{eq:NS} and \eqref{eq:mainresult} for the $1/\ell$ correction to $\partial_t\langle\mathcal{q}_2(x,t)\rangle$, computed using the analytical solutions outlined in \cite{SM}.
The results from Eq. \eqref{eq:mainresult} are in excellent agreement with the numerical data, being always within $2$ times the error-bars, as is shown in the inset. Meanwhile, the predictions by Kubo-like (NS equation) diffusion deviate from the numerical data in a statistically significant way, sometimes more than $5$ error-bars.

\prlsection{Discussion}
We introduced a new theory for diffusion in systems with linearly degenerate ballistic modes, particularly applicable to integrable models. Unlike the NS equation, which describes the averages of charge-densities, our theory consists of two coupled equations for both the averages and correlations of charge-densities. This approach reveals that the Kubo-formula based diffusion fails at late times due to the violation of the local equilibrium assumption by correlations.
We numerically verified our theory in the hard-rod gas and found it consistent with independent derivations~\cite{UPCOMING}. Note that in integrable systems, the Onsager matrix was evaluated exactly for the first time in \cite{PhysRevLett.121.160603}. Our linear-response results agree with this Onsager matrix, but the claim made there for the full diffusive-scale hydrodynamic equation, which was based on Kubo-like diffusion, is only a good approximation at early times.

Future directions include extending these results to higher orders, incorporating external potentials, and exploring the nature of thermalization. Unlike NS-like diffusion, which leads to thermalization due to entropy increase, our theory lacks a clear increasing entropy function. Identifying such an entropy for our equations and classifying maximum entropy states would be valuable.

This work also raises general questions about diffusion in hydrodynamics beyond linearly degenerate systems. The NS equation, crucial in physics, does not account for correlations, which are essential in our new theory. It seems plausible that correlations should influence the dynamics of generic linearly non-degenerate fluids. Investigating their impact and implications is an interesting avenue for future research.

\begin{acknowledgments}
\prlsection{Acknowledgments} We acknowledge inspiring discussions with Romain Vasseur, Sarang Gopalakrishnan and Takato Yoshimura. J.D.N. and L.B. acknowledge discussion on related topic with Maciej Łebek and Miłosz Panfil. B.D. is also grateful to Olalla Castro Alvaredo, David Horvath and Paola Ruggiero for related discussions. B.D. is particularly thankful to Paola Ruggiero and Tony Jin for discussions in 2022 at King's College London, where they expressed the general idea of a state generalising GGEs by including bilocal charges, which was the inspiration for \eqref{eq:nonlocalgge}. J.D.N. and L.B. are funded by the ERC Starting Grant 101042293 (HEPIQ) and the ANR-22-CPJ1-0021-01. FH acknowledges funding from the faculty of Natural, Mathematical \& Engineering Sciences at King’s College London. BD was supported by the Engineering and Physical Sciences Research Council (EPSRC) under grant EP/W010194/1.
\end{acknowledgments}

\bibliography{apssamp}

\clearpage
\begin{center}
{\Large End Matter -\titleinfo
}
\end{center}

\prlsection{Regularization of observables}
Assuming \eqref{eq:BMFT} to be accurate up to $1/\ell$, needs particular care. Indeed, delta-correlations $S_{ij} \sim \ell^{-1}C_{ij}\delta(x-y)$ would make any non-linear function of $q_i(x)$'s at a given point $x$ (keeping $t$ implicit) simply ill-defined. Physically, random variables should, in fact, be functions of fluid-cell averages, $\ep^{-1} \int _{-\ep/2}^{\ep/2}\dd y\,q(x+y)$, for macroscopic size $\ep$ as small as possible. Such functions are well-defined, but delta-correlations give unphysical dependence on the fluid-cell size $\ep$. However, these delta-correlations represent fast, microscopic-scale fluctuations which are already encoded within the Euler-scale functions $\mathcal a(q_\cdot)$ used to represent the random variables. Indeed, these functions are averages of the corresponding microscopic observable within maximal entropy states, and such averages involve microscopic fluctuations. Thus, the random variables $\mathcal a(q_\cdot)$ should only be affected by slow, macroscopic-scale fluctuations: those that are of order $1/\ell$. This justifies our main proposal: random variables representing local observables should be regularised as $[\mathcal a(q_\cdot(x))]^{\rm reg}$, where we perform fluid-cell averaging by {\em avoiding equal coordinates}. This, intuitively, only considers potential large-scale fluctuations, which, by the BMFT, are encoded within long-range correlations. That is, we define, for any product of conserved densities, its regularised expression
\beq\label{eq:regularisation}\begin{aligned}
	&[q_{i_1}(x)\cdots q_{i_n}(x)]^{\rm reg}\\
	&=\lim_{\ep\to0}
	\frc1{\ep^n}
	\int_{\Lambda_\ep}
	\dd y_1\cdots \dd y_n\,
	q_{i_1}(x+y_1)\cdots q_{i_n}(x+y_n),
	\end{aligned}
\eeq
where $\Lambda_\ep$ is $[x-\ep/2,x+\ep/2]^n$ with the condition $y_i\neq y_j\forall i\neq j$, and write
\beq\label{eq:proposal}
	\bra a(x,t)\ket = \dbra [\mathcal a(q_\cdot(x,t))]^{\rm reg}\dket,
\eeq
and similarly for higher-point correlation functions.  

\begin{figure}[h]
\includegraphics[width=\linewidth]{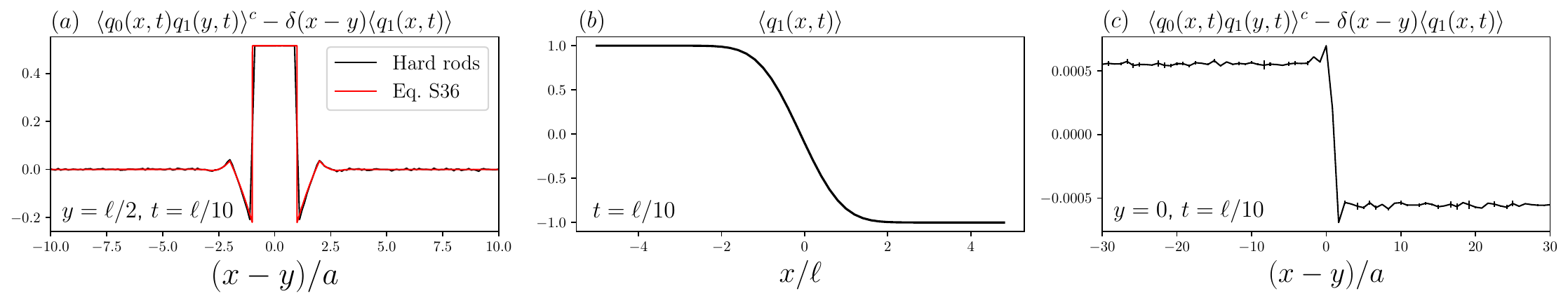}
\caption{(a) Plot for the microscopic correlations in a Hard rods gas, with rods length $a=0.3$ and scale $\ell=200$. Respectively, from left to right, we plot $\langle q_0(x,t)q_1(y,t)\rangle^c_{\rm reg}-\delta(x-y)\langle q_1(x,t)\rangle$
at the points $y=\ell/2$, $t=\ell/10$ and $\ell=200$. 
The Hard rods data (black line) is compared with prediction from Eq. \eqref{eq:paircorrHR1} (red line).
The Hard rod data are averaged over $10^7$ initial states. Eq. \eqref{eq:paircorrHR1} has been evaluated truncating the summation at $k=50$, such that the truncation error is $<10^{-16}$.
The agreement between the prediction and the numerical data is excellent, up to the Hard rods monte carlo noise. We stress that discrepancies induced by long-range correlations are expected to be order $\mathcal{O}(\ell^{-1})$. (b) Plot of $\langle q_1(x,t)\rangle$ at a time $t=\ell/10$ from Hard rods numerics.
The figure (c) shows the microscopic correlations $\langle q_0(x,t)q_1(y,t)\rangle^c_{\rm reg}-\delta(x-y)\langle q_1(x,t)\rangle$ at $y=0$ and $t=10/\ell$. At this point, the leading contribution in $\ell$ is expected to be vanishing, having $\langle q_0(x=0,t=\ell/10)\rangle\simeq 0$ (as shown in box (b)), and since $\langle q_0(x,t)q_1(y,t)\rangle^c_{\rm reg}\rangle\propto\langle q_1(x,t)\rangle+\mathcal{O}(\ell^{-1})$.
Hence, this plots shows the microscopic structure of the discontinuity in the two-point correlation. We can conclude that the 'jump' is developed in a length scale $\sim a$.}
\label{fig: SM Micro corr}
\end{figure}

Clearly, because $t=0$ correlations in local GGEs are zero at unequal points (there are no large-scale fluctuations), under this prescription $\dbra [\mathcal a(q_\cdot(x,0))]^{\rm reg}\dket = \mathcal a(\mathcal q_\cdot(x,0))$, as required. 
\\
\\
\prlsection{Validity of the NS equation at early times and in linear response}
Starting from a local GGE \eqref{eq:hydro01}, the early times solutions of our new hydrodynamic equations \eqref{eq:mainresult}, \eqref{eq:timeevCorr} {\em approximately satisfy the Navier-Stoker equation \eqref{eq:NS}}. Let us solve the BMFT equation \eqref{eq:timeevCorr} for long-range correlations explicitly, at early times, with initial condition $E(x,y;0)=0$. Normal modes $n_j = R_j^{~k} q_k$, diagonalise the flux Jacobian \(A = R v^{\rm eff} R^{-1}\), where the diagonal matrix \(v^{\rm eff}\) contains the modes velocities. One may choose the matrix $R$ such that \cite{Spohn2014} $R C R^{-1}= \mathbf{1}$. By \eqref{eq:shapeofCorr} normal-mode correlations take the form $\langle n_k(x,t) n_l(y,t) \rangle^{\rm c}= \delta(x-y){\bf 1} + \langle n_k(x,t) n_l(y,t) \rangle^{\rm c}_{\rm lr}$ where the second term is regular. At early times, we only require the delta-terms in \eqref{eq:timeevCorr} to agree. These are accounted for by jumps, so we make the following ansatz:
\begin{align}\label{ansatz2text}
    & \langle n_i(x,t) n_j(y,t) \rangle^{\rm c}_{\rm lr} = \frac{1}{2} \Delta_{ij}(x,y) \sgn(x-y) \\
    & \qquad + \frac{1}{2} K_{ij}(x,y) \sgn(x-y -(v_i^{\rm eff}t - v_j^{\rm eff}t)),\quad \mbox{($t$ small)}\no
\end{align}
where $\Delta$ and $K$ are continuous in $x,y$. Applying $x$ any $y$ derivatives on the sign function in the first term on the right-hand side gives the delta function on the right-hand side of \eqref{eq:timeevCorr}, and thus we require
\beq\label{eq:delta}
	\Delta_{ij}(x,x) = -\frc{\bra n_i,n_j,j^-_k\ket C^{kr}\p \mathcal q_r}{v_i^{\rm eff} - v_j^{\rm eff}}.
\eeq
In fact, \eqref{eq:delta} holds for all times: \textit{long-range correlation functions display jumps} at equal positions, proportional to the local spatial variation of the densities, and fully determined by the local state. Our numerical analysis shows that the width of the jump is of microscopic scale (see SM \cite{SM}, Fig. \ref{fig: SM Micro corr}). Eq.~\eqref{eq:delta} makes sense if $\bra n_i,n_j,j^-_k\ket=0$ whenever $v_i^{\rm eff} = v_j^{\rm eff}$: this property of {\em three-point coupling degeneracy} holds in linearly degenerate systems, as we show \cite{SM} using hydrodynamic projections \cite{doyonemergenceeuler,ampelogiannisdoyon} and non-linear response \cite{10.21468/SciPostPhys.5.5.054,Doyon2022}.

The sign function on the second term in \eqref{ansatz2text} annihilates the left-hand side of \eqref{eq:timeevCorr} at early times as the rotation matrix $R$ may be considered constant, and normal modes evolve with velocities $v_i^{\rm eff}$. We thus see that a front opens up, producing the jump at equal coordinates; the local GGE is unstable, as discontinuities in long-range correlations immediately appear. The initial condition requires $\Delta = - K$, and therefore $E^{\rm sym}_{ij} = \sgn(v_i^{\rm eff}-v_j^{\rm eff})
\frc{\Delta_{ij}}2$. Putting this together with \eqref{eq:delta}, we find the expression of the diffusion matrix $\mathfrak{D}_i^{~s}$ in terms of quadratic-charge projection first obtained in \cite{Doyon2022} and known to agree with the Kubo formula in integrable systems
\begin{align}\label{eq:earlytime}
   \lim_{t \to 0^+}   &   \frac{\delta   \mathcal{j}_i(q_{\cdot})   }{{\delta \mathcal{q}_r \delta \mathcal{q}_k}} E^{\rm sym}_{rk}  \\&  = -  \frac{\langle j_i^- , n_r, n_l \rangle  \langle j_k^- , n_r, n_l \rangle   C^{ks}} {|v_r^{\rm eff} - v_l^{\rm eff}|} \partial_x \mathcal{q}_s = -  \mathfrak{D}_i^{~s} \partial_x \mathcal{q}_s.\nonumber
\end{align}
At nonzero times  \(\Delta \) and \(K\) satisfy different evolution equations, thus \(\Delta\neq -K\) and corrections to \eqref{eq:earlytime} accumulate.

By a similar argument, {\em at linear order, our new diffusive-scale equation reproduces the linearised NS equation}, $\partial_t \bra q_i(x,t)q_r(0,0)\ket^{\rm c} + A_{i}^{~k} \partial_x \bra q_k(x,t)q_r(0,0)\ket^{\rm c} = \frac{1}{2} \mathfrak{D}_i^{~k} \partial_x^2 \bra q_k(x,t)q_r(0,0)\ket^{\rm c}$, within GGEs. This can be done by performing linear response theory on the long-range GGE \eqref{eq:nonlocalgge}: $\beta\to\beta+\delta\beta$ and $\gamma\to0+\delta\gamma$, where we note that in a homogeneous GGE, we have $\gamma=0$. As the rotation matrix $R$ can be taken constant, the ansatz \eqref{ansatz2text} gives an exact solution (see \cite{SM} for the explicit form of $\Delta(x,y)$ and $K(x,y)$). Hence, Eq.~\eqref{eq:earlytime} holds at all times.
\\
\\
\prlsection{Microscopic correlations in the hard rod gas}
\label{sec:SM_HR_microscopic_corr}
In this section, we show numerical results for the microscopic correlation functions in a hard rods gas.
Let us consider a homogeneous hard rods gas, with rods' length $a$ and particle density $\rho(\theta)=\bar\rho h(\theta)$, with $\int\dd{\theta}h(\theta)=1$ and $\theta$ the particles' velocities.
The microscopic correlation function
for such a system is given by \cite{Flicker68}
\begin{multline}
\label{eq:paircorrHR1}
    S^{\rm micro}_{\theta,\theta'}(x,x')= \delta(x-x')\delta(\theta-\theta')\bar\rho h(\theta) +\\+(\mathfrak{n}^{(2)}(x-x')-\bar\rho^2) h(\theta)h(\theta').
\end{multline}
\begin{multline}
\label{eq:paircorrHR2}
    \mathfrak{n}^{(2)}(x)=\frac{\bar\rho^2}{1-\bar\rho a} \sum_{k=1}^\infty \frac{1}{(k-1)!} \left( \frac{|x|-ak}{\bar\rho^{-1}-a}\right)^{k-1} \times\\\times \exp \bigg[] - \left( \frac{|x|-ak}{\bar\rho^{-1}-a} \right)\bigg] \Theta_H(|x|-ka)\,,
\end{multline}
where $\Theta_H$ is the Heaviside step function.
Let us now consider a non-uniform gas with $\ell$ the typical scale of spatial variations. Its correlations will be given by $\langle\rho(x,\theta)\rho(x,\theta')\rangle^c=S^{\rm micro}_{\theta,\theta'}(x,x')+\mathcal{O}(\ell^{-1})$, where $S^{\rm micro}_{\theta,\theta'}(x,x')$ is defined by taking $\bar\rho\to\bar\rho(x,t)$ with $\bar\rho(x,t)\equiv \int\dd{\theta}\rho(x,\theta,t)$ and $h(\theta)\to h_x(\theta,t)\equiv\rho(x,\theta,t)/\bar\rho(x,t)$. We also stress that here $(x,x')$ are microscopic spatial coordinates.

We compare the numerical results for the microscopic correlations functions of a time evolving hard rod gas with the prediction of Eq. \eqref{eq:paircorrHR1}.
More precisely, we consider a gas of hard rods with only two modes, $\theta_+=+1$ and $\theta_-=-1$, evolving from the initial state $\rho_{\pm}(x,t=0)=1\mp{\rm Erf}(x/\ell)/2$. In the initial state also $q_0(x,t=0)=\rho_{+}(x,t=0)+\rho_{-}(x,t=0)=1$ and $q_1(x,t=0)=\rho_{+}(x,t=0)-\rho_{-}(x,t=0)=-{\rm Erf}(x/\ell)$.

In Fig. \ref{fig: SM Micro corr}(a) we show the hard-rod numerical results for $\langle q_0(x,t)q_1(y,t)\rangle^c-\delta(x-y)\langle q_1(x,t)\rangle$
at the points $y=\ell/2$ and $t=\ell/10$ for $\ell=200$. Comparing it with the theoretical predictions fro Eq. \eqref{eq:paircorrHR1}, we observe excellent agreement. We also stress that all the discrepancies induced by the long-range correlations are at order $\mathcal{O}(\ell^{-1})$, hence they are not visible in Fig. \ref{fig: SM Micro corr} (a).
More precisely, we used the relation
\begin{equation}
\begin{split}
    \langle q_0(x)&q_1(y)\rangle^c=\sum_{i,j=1}^2  S^{\rm micro}_{\theta_i,\theta_j}(x,y)\theta_i\theta_j+\mathcal{O}(\ell^{-1}) \\&\mbox{with:}\quad \sum_{i,j=1}^2  S^{\rm micro}_{\theta_i,\theta_j}(x,y)\theta_i\theta_j\propto\langle q_1(y)\rangle
\end{split}
\end{equation}
where we used $\theta_i=[-1,+1]$ and where $S^{\rm micro}$ is defined in Eq. \eqref{eq:paircorrHR1}.
But, as it is possible to observe from Fig. \ref{fig: SM Micro corr} (b), $\langle q_1(y,t)\rangle$ is vanishing at $t=\ell/10$ and $y=0$. Thus, we expect the leading contribution to be vanishing at this point. This permits to isolate the long-range contribution to the correlations at microscopic scale and to observe the microscopic structure of the discontinuity introduced in the main text. In particular, in Fig. \ref{fig: SM Micro corr} (c), we show the two point function $\langle q_0(x,t)q_1(y,t)\rangle^c-\delta(x-y)\langle q_1(x,t)\rangle$ at $t=\ell/10$ and $y=0$. From this picture, we can observe that the discontinuity predicted at hydrodynamic scale, is developed in the gas at scales $\sim a$.
\onecolumngrid

\twocolumngrid

\onecolumngrid
\newpage 
\appendix

\begin{center}
{\Large Supplementary Material -  
\titleinfo
}
\end{center}

\setcounter{equation}{0}
\setcounter{figure}{0}
\setcounter{table}{0}
\setcounter{page}{1}
\renewcommand{\theequation}{S\arabic{equation}}
\setcounter{figure}{0}
\renewcommand{\thefigure}{S\arabic{figure}}
\renewcommand{\thepage}{S\arabic{page}}
\renewcommand{\thesection}{S\arabic{section}}
\renewcommand{\thetable}{S\arabic{table}}
\makeatletter

\renewcommand{\thesection}{\arabic{section}}
\renewcommand{\thesubsection}{\thesection.\arabic{subsection}}
\renewcommand{\thesubsubsection}{\thesubsection.\arabic{subsubsection}}

\section{Dynamics of two-point correlations }

In this section, we derive the evolution equation for the regular part of the correlations. 
According to the BMFT, or alternatively the continuity equation combined with the projection principles, the correlation functions satisfy the evolution equation~\cite{10.21468/SciPostPhys.5.5.054,10.21468/SciPostPhys.15.4.136}
\begin{equation} 
\partial_t S_{ij} + \partial_x \big(A_i^{~k} S_{kj}\big) + \partial_y \big(S_{ik} A_j^{~k}\big) = 0.
\end{equation}
Now, substituting the generic form for the correlations in terms of diverging part plus the regular part,
\begin{equation}\label{qiqjlr}
    S_{ij}(x,y,t) = \delta(x-y) C_{ij}(x,t) + E_{ij}(x,y,t),
\end{equation}
we aim at writing the evolution equation for the $E_{ij}(x,y,t)$. We look at (a) the $\delta'(x-y)$, (b) the $\delta(x-y)$ part coming from the $\delta(x-y)C_{ij}(x,t)$ part.

(a) This reads
\beq
   S_{im } A_j^{m}(y,t) \stackrel{\delta}=
    C_{im}(x,t)\delta(x-y) A_j^{~m}(x,t).
\eeq
So we have 
\beq
    \delta'(x-y)\times \big(
    AC - CA^T\big)_{ij} = 0,
\eeq
where we use matrix notation, and we use the symmetry of the $B_{ij} = A_i^{~k}C_{kj}$ matrix~\cite{Doyon2020}.

(b) The term proportional to $\delta$ is given by
\beq\label{directdelta}
    \delta(x-y)\times \Big(
    \p_t C_{ij} + \p_x B_{ij}\Big).
\eeq
This is evaluated as
\beq
    \p_t C_{ij} = \bra q_i,q_j,q_k\ket\,
    C^{kr}\p_t \bra q_r\ket
    = -\bra q_i,q_j,q_k\ket\,
    C^{kr}A_r^{~s}\p_x \bra q_s\ket
    = -\bra q_i,q_j,q_k\ket\,
    A_r^{~k}C^{rs}\p_x \bra q_s\ket,
\eeq
where we used $C^{-1}A = A^TC^{-1}$, and
\beq
    \p_x B_{ij} = \bra j_i,q_j,q_k\ket\,
    C^{kr}\p_x \bra q_r\ket
    = \bra q_i,q_j,j_k\ket\,
    C^{kr}\p_x \bra q_r\ket,
\eeq
where we also used complete symmetry of the current-density-density 3-point coupling. Therefore
\beq\label{resptC}
    \p_t C_{ij} + \p_x B_{ij}
    = \bra q_i,q_j,j^-_k\ket C^{kr}\p_x \bra q_r\ket.
\eeq
We then obtain the equation reported in the main text: 
\begin{equation}
    \p_t E_{ij} + \p_x \big(A_i^{~k}E_{kj}\big)
    +\p_y \big(D_{ik}A_j^{~k}\big)
    = \bra q_i,q_j,j_k^-\ket \p_x \beta^k \delta(x-y).
\end{equation}
The latter can be conveniently recast in the normal mode basis $E_{ij}(x,y;t) = (R^{-1})_i^{~k}(x,t)\bra n_k(x,t)n_l(y,t)\ket^{\rm c}_{\rm lr} (R^{-1})_j^{~l}(y,t)$,
\begin{equation}\label{equ:sm_dyntwopoint_E_eq}
    \p_t (R^{-1} \bra nn\ket^{\rm c}_{\rm lr} R^{-T})
    +
    \p_x \big( R^{-1} v^{\rm eff}\bra nn\ket^{\rm c}_{\rm lr}R^{-T}\big)
    + \p_y \big( R^{-1}\bra nn\ket^{\rm c}_{\rm lr}v^{\rm eff} R^{-T}\big) =
    \bra q_i,q_j,j_k^-\ket \p_x \beta^k \delta(x-y).
\end{equation}
This equation is solved by the following ansatz:
\beq\label{ansatz2}
    \bra n_i(x,t)n_j(y,t)\ket^{\rm c}_{\rm lr}
    =
    \Delta_{ij}(x,y;t) \,\frc12 \sgn(x-y) + K_{ij}(x,y;t)\,\frc12\sgn(u_i(x,t)-u_j(y,t)),
\eeq
where $u(x,t)$ are the characteristics functions, satisfying
\begin{equation}
    \p_t u_i(x,t) + v^{\rm eff}_i(x,t)\p_x u_i(x,t) = 0,\quad u_i(x,0) = x.
\end{equation}
The quantity $u_i(x,t)$ is the position at time 0 from which the characteristics $i$ get to position $x$ at a time $t$. The initial condition imposes
\begin{equation}
    \Delta_{ij}(x,y;0) = - K_{ij}(x,y;0).
\end{equation}
Since the $\delta(x-y)$ on the rhs of \eqref{equ:sm_dyntwopoint_E_eq} can only be compensated by a spatial derivative on $\sgn(x-y)$ the local jump is fixed by:
\beq
    (v_i^{\rm eff}(x,t) - v_j^{\rm eff}(x,t))\Delta_{ij}(x,x;t) = \bra n_i,n_j,j^-_k\ket(x,t)\, \p_x\beta^k(x,t).
\eeq

At this point, we can see that this does not have a solution unless $\bra n_i,n_i,j^-_k\ket(x,t) = 0$. In the next section we show that this condition follows from linear degeneracy. Note that this is the main reason why our ansatz only applies to linearly degenerate systems. Assuming linear degeneracy, we find:
\beq\label{DeltaSgeneral}
    \Delta_{ij}(x,x;t) = \frc{\bra n_i,n_j,j^-_k\ket(x,t)\, \p_x\beta^k(x,t)} {v_i^{\rm eff}(x,t) - v_j^{\rm eff}(x,t)}.
\eeq

For the remaining terms, the function $\sgn(u_i(x,t)-u_j(y,t))$ automatically satisfies the evolution equation, so we can divide the rest into two equations:
\begin{equation}
    \p_t (R^{-1} \Delta S R^{-T})
    + \p_x (R^{-1} v^{\rm eff} \Delta R^{-T}) + \p_y (R^{-1} \Delta v^{\rm eff}  R^{-T}) = 0
\end{equation}
and (in a reduced notation)
\begin{equation}
    \p_t (R^{-1} K\widehat{\sgn(u-u)} R^{-T})
    + \p_x (R^{-1} v^{\rm eff} K\widehat{\sgn(u-u)}R^{-T}) + \p_y (R^{-1} K\widehat{\sgn(u-u)} v^{\rm eff}  R^{-T})  = 0.
\end{equation}
where $\widehat{\sgn(u-u)}_{ij}$ is $\sgn(u_i(x,t)-u_j(y,t))$ and the hat means that the derivative does not apply on it. Because derivatives apply on $R^{-1}$ and $R^{-T}$, and because of the matrix structure of $\sgn(u-u)$, the resulting equation for $\Delta$ and for $K$ are different; hence the relation $\Delta = -K$ is broken at non-zero times. This is (technically) why the standard diffusion formula breaks at later times. But it is recovered at the linear order for perturbations on top of a constant background, where we can take $R$ to be a constant: in this case both $\Delta$ and $K$ satisfy $\p_t \Delta + \p_x (v^{\rm eff} \Delta) + \p_y (\Delta v^{\rm eff}) = 0$, and hence the relation $\Delta = -K$ holds for all times.

Plugging the formula $\Delta = - K$ into rhs of \eqref{eq:mainresult} we obtain the following result:
\begin{align}
    \frac{\delta   \mathcal{j}_i(\mathcal q_{\cdot})   }{{\delta \mathcal{q}_r \delta \mathcal{q}_k}} \,E^{\rm sym}_{rk} &= \frac{\delta   \mathcal{j}_i(\mathcal q_{\cdot})   }{{\delta \mathcal{q}_r \delta \mathcal{q}_k}} \,K_{rk}(x,x;t)\,\frac{\sgn(u_r(x^-,t)-u_k(x^+,t))+\sgn(u_r(x^+,t)-u_k(x^-,t))}{2}\\
    &=\frac{\delta   \mathcal{j}_i(\mathcal q_{\cdot})   }{{\delta \mathcal{q}_r \delta \mathcal{q}_k}} \,K_{rk}(x,x;t)\,\sgn(v^{\rm eff}_k-v^{\rm eff}_r)\\
    &=-\langle q_a,q_b,j_i^-\rangle(x,t) C^{ar}C^{bk} (R^{-1})_r^{~c}(R^{-1})_k^{~d}\,\frc{\bra n_c,n_d,j^-_k\ket(x,t)\, C^{kl}\p_xq_l(x,t)} {v_r^{\rm eff}(x,t) - v_k^{\rm eff}(x,t)}\,\sgn(v^{\rm eff}_k-v^{\rm eff}_r)\\
    &=\frc{\langle n_a,n_b,j_i^-\rangle(x,t) \bra n_r,n_k,j^-_k\ket(x,t)\, C^{kl}\p_xq_l(x,t)} {|v_r^{\rm eff}(x,t) - v_k^{\rm eff}(x,t)|} = -\mathfrak{D}_{i}^{~l}\p_xq_l(x,t).\label{eq:sm_early_times}
\end{align}
This is the result of the quadratic charge projection Kubo formula in integrable models~\cite{Doyon2022}, which therefore still holds for short times and weak perturbations.

\section{Degenerate three-point coupling from linear degeneracy}

In this section we show that the property of {\em degenerate three-point coupling},
\beq\label{deg3point}
    \bra n_i,n_j,a^-\ket=0 \quad \mbox{if $v_i^{\rm eff} = v_j^{\rm eff}$, for any local observable $a$},
\eeq
follows from the property of {\em linear degeneracy}
\beq\label{lindegsm}
    \frc{\p v^{\rm eff}_i}{\p n_i} = 0.
\eeq
Recall that $n_i$'s are normal modes, and that we denote by $\p n_i/\p q_j = R_i^{~j}$ the transformation Jacobian.

First, we mention that there appears to be no direct link between \eqref{lindegsm} and \eqref{deg3point}, using solely the basic properties of Euler hydrodynamics ``calculus". For simplicity we restrict to the strictly hyperbolic case where no two $v^{\rm eff}_i$ are equal, and we consider the observable $a = R_l^{~k}j_k$. Observe:
\begin{align}
    R_l^{~k}\langle n_i,n_j,j^-_k\rangle &= R_l^{~k}R_i^{~a}R_j^{~b}\langle q_a,q_b,j^-_k\rangle = R_l^{~k}R_i^{~a}R_j^{~b}C_{ac}C_{bd}\tfrac{\p}{\p q_d}A_k^{~c}\\
    &= R_l^{~k}R_i^{~a}R_j^{~b}C_{ac}C_{bd}\tfrac{\p}{\p q_d}((R^{-1})_{k}^{~e}v^{\rm eff}_e R_{e}^{~c})\\
    &= R_l^{~k}R_i^{~a}C_{ac}\tfrac{\p}{\p n_j}((R^{-1})_{k}^{~e}v^{\rm eff}_e R_{e}^{~c})\\
    &= (R_l^{~k}\tfrac{\p}{\p n_j}(R^{-1})_{k}^{~i})v^{\rm eff}_i + \delta_{il}\tfrac{\p}{\p n_j}v^{\rm eff}_l + v^{\rm eff}_l(R^{-1})_{ic}\tfrac{\p}{\p n_j}(R_{l}^{~c})\\
    &= (v^{\rm eff}_i-v^{\rm eff}_l) (R_l^{~k}\tfrac{\p}{\p n_j}(R^{-1})_{k}^{~i})  + \delta_{il}\tfrac{\p}{\p n_j}v^{\rm eff}_l
\end{align}
In the last line, we used $RCR^T = \mathbf{1}$ (The normal modes can always be chosen in a way that his condition holds~\cite{Spohn2014}). We require that this vanishes if $i=j$ (or more generally if $v^{\rm eff}_i = v^{\rm eff}_j$) for any $l$. If linear degeneracy holds, the second term vanishes. However, linear degeneracy does not fix the first term. Note that by setting $i=l$ the first term vanishes, and thus we require a stronger condition than linear degeneracy.

The proof instead requires the use of an  additional input from the general properties of correlation functions; thus degenerate 3-point coupling is an additional condition to impose on the formal structure of Euler hydrodynamics. The input is the vanishing
\beq
    \lim_{T\to\infty} T^{-1}\int_0^T \dd t\int \dd x\,\bra a(x,0)b^-(0,t)\ket^{\rm c} = 0
\eeq
for any observable $a,b$ supported on finite regions of space. This holds by projection mechanisms, and was shown rigorously in every short-range quantum spin chains and lattices in \cite{doyonemergenceeuler,ampelogiannisdoyon}.

As mentioned, by linear degeneracy, solutions to the Euler equation are unique and do not develop shocks \cite{FERAPONTOV1991112,lindeg1,lindeg2,hübner2024newquadraturegeneralizedhydrodynamics}. Therefore, non-linear response theory as proposed in \cite{10.21468/SciPostPhys.5.5.054} can be applied: observables at time 0 can be inserted within correlation functions of any order, by functional differentiation of the initial state. One result of this, as shown in \cite[App E]{Doyon2022}, is the expression of 3-point functions
\beq
    \bra n_i (x,0) n_j(y,0) a^-(z,t)\ket^{\rm c} = \bra n_i,n_j,a^-\ket
    \delta(z-x-v^{\rm eff}_i t)\delta(z-y - v^{\rm eff}_j t).
\eeq
Choosing $n_i, n_j$ to have zero expectation value (by appropriate shifting by the identity observable), and performing the integral $\int_{x-\ep}^{x+\ep} \dd y$, the left-hand side is
\beq
    \bra O_\ep(x,0) a^-(z,t)\ket^{\rm c}
\eeq
where
\beq
    O_\ep(x,0) = \int_{x-\ep}^{x+\ep} \dd y\,n_i(x,0)n_j(y,0)
\eeq
is an observable supported on a finite region of space of length $2\ep$. The right-hand side, on the other hand, gives
\beq
    \bra n_i,n_j,a^-\ket
    \delta(z-x-v^{\rm eff}_i t)
    \chi(|v^{\rm eff}_i  - v^{\rm eff}_j| t < \ep).
\eeq
We must therefore have the equality of
\beq
    \lim_{T\to\infty} T^{-1}\int_0^T \dd t\int \dd x\,\bra O_\ep(x,0) a^-(z,t)\ket^{\rm c} = 0
\eeq
with
\beq
    \lim_{T\to\infty} T^{-1}\int_0^T \dd t\int \dd x\,
    \bra n_i,n_j,a^-\ket
    \delta(z-x-v^{\rm eff}_i t)
    \chi(|v^{\rm eff}_i  - v^{\rm eff}_j| t < \ep)
    = \bra n_i,n_j,a^-\ket\,
    \delta_{v^{\rm eff}_i,v^{\rm eff}_j}
\eeq
which shows \eqref{deg3point}.

\noindent Remark: Plugging \eqref{deg3point} into \eqref{eq:sm_early_times} we see that linear degeneracy is required for a finite diffusion matrix $\mathfrak{D}_i^{~k}$ and the Onsager matrix $\mathfrak{L}_{ij}$.

\section{Long-Range equilibrium states}

We here consider the non-local, long-range equilibrium state introduced in the main text 
\begin{align}\label{eq:nonlocalgge-SM}
	\rho_{\rm LR} \propto &\exp \Big[-\int \dd x\, \beta^i(x) q_i(x) +  \int_{x\neq y} \hspace{-0.3cm}\dd x \dd y\, \frc{\gamma^{ij}(x,y)}2 \delta {q}_i(x)  \delta {q}_j(y) \Big] ,
\end{align}
with the centered fluctuation $\delta {q}_i = q_i - \langle q_i \rangle$
and we shall show that the functions $\gamma^{ij}$ fix the regular part of the correlations, equivalently to what the chemical potentials do for the charges,  as
\beq\label{eq:finalres1}
    E_{kl}(0,z)
    =
    2\ell^{-1}
	\gamma^{ij}(0,z)
    C_{ik}(0) C_{jl}(z)
    +O(\ell^{-2}),
\eeq
which indeed gives the correct cumulant expansion for the local operators in terms of the correlations 
\beq\label{eq:finalres2}
    \bra a(0)\ket_{\rm LR}
    = a(\bra q(0)\ket_{\rm LR})
    +\frc12 \ell^{-1} \frc{\p^2}{\p q_i \p q_j}\bra a\ket_{\bra q(0)\ket}
    E^{\rm sym}_{ij}(0^+,0^-)
\eeq
in agreement with the point-splitting regularisation introduced in the main text.

In order to show this, we consider the average of a local one or two-point observable that be obtained, to first order is $\ell^{-1}$, by expanding the non-local term
\beqa
	\bra a(0)\ket_{\rm LR} &=& \bra a(0)\ket
	+ \ell
	\int \dd x\dd y\,\gamma^{ij}(x,y)
	\bra (q_i(x)-\bra q_i(x)\ket)(q_j(y)-\bra q_j(y)\ket),a(0)\ket + \ldots \n
	&=& \bra a(0)\ket
	+
	\ell\int \dd x\dd y\,\gamma^{ij}(x,y)
	\bra q_i(x)q_j(y)a(0)\ket^{\rm c}
    + \ldots \n
\eeqa
We now shall use some hypothesis on the structure of the three-point functions 
\beq
    \bra q_i(x)q_j(y)a(0)\ket^{\rm c} = 0
    \ \ (|x|+|y|>r_{\rm micro}\ell^{-1}),
\eeq
which clearly implies the following relations 
\beq
    \int \dd x\dd y\,\bra q_i(x)q_j(y)a(0)\ket^{\rm c}
    = \ell^{-2}\frc{\p^2}{\p\beta^i\p\beta^j} \bra a\ket_{\bra q(0)\ket}+ O(\ell^{-3}).
\eeq
We note that, by using PT symmetry, if $a(x)$ is PT symmetric then
\beq\label{PTsymqqa}
    \bra q_i(x)q_j(y)a(0)\ket^{\rm c}
    =
    \bra q_i(-x)q_j(-y)a(0)\ket^{\rm c},
\eeq
which implies 
\beq
    \ell\int \dd x\dd y\,\gamma^{ij}(x,y)\bra q_i(x)q_j(y)a(0)\ket^{\rm c} =
    \ell^{-1}\gamma^{ij}_{\rm s}(0^+,0^-)\frc{\p^2}{\p\beta^i\p\beta^j} \bra a\ket_{\bra q(0)\ket}
    + O(\ell^{-2})
\eeq
where we wrote $\gamma^{ij}(x,y) = \gamma^{ij}_{\rm s}(x,y) + \gamma^{ij}_{\rm as}(x,y)$
\beq
    \gamma^{ij}_{\rm s}(x,y) = \frc{\gamma^{ij}(x,y) + \gamma^{ij}(-x,-y)}{2},\quad
    \gamma^{ij}_{\rm as}(x,y) = \frc{\gamma^{ij}(x,y) - \gamma^{ij}(-x,-y)}{2}
\eeq
and the anti-symmetric part cancels by \eqref{PTsymqqa}. For the symmetric part, we wrote $\gamma^{ij}_{\rm s}(x,y) = \gamma^{ij}_{\rm s}(0^+,0^-) + O(\ell^{-1})$ for any $x,y = O(\ell^{-1})$; in particular, although $\gamma^{ij}_{\rm s}(x,y)$ is not continuous at $x=y$, the discontinuity is small, $\Delta\gamma^{ij}_{\rm s}(x) := \gamma^{ij}_{\rm s}(x+0^+,x-0^+) - \gamma^{ij}_{\rm s}(x-0^+,x+0^+) = (\Delta\gamma^{ij}(x) - \Delta\gamma^{ij}(-x))/2 = O(\ell^{-1})$ if $x=O(\ell^{-1})$ (using smoothness along the tubular neighbourhood of the diagonal). Hence we obtain  
\beq
    \bra a\ket_{\rm LR} = a(\bra q(0)\ket)
    + \ell^{-1}\gamma^{ij}_{\rm s}(0^+,0^-)
    \frc{\p^2}{\p\beta^i\p\beta^j}\bra a\ket_{\bra q(0)\ket} + O(\ell^{-2}).
\eeq
We can now express everything in terms of the charges contained in the $\rm LR$ state. Writing things together, and using $\bra a(0)\ket = a(\bra q(0)\ket)$ (which is granted by the PT symmetry of $a(x)$), we find
\beq
    \bra a(0)\ket_{\rm LR}
    = a(\bra q(0)\ket_{\rm LR}) +
    \ell^{-1}\gamma_{\rm s}^{ij}(0^+,0^-)\Big(
    \frc{\p^2}{\p\beta^i\p\beta^j} \bra a\ket_{\bra q(0)\ket}
    - \frc{\p^2}{\p\beta^i\p\beta^j} \bra q_l\ket_{\bra q(0)\ket} \frc{\p }{\p q_l} \bra a\ket_{\bra q(0)\ket}\Big)
    +O(\ell^{-2})
\eeq
which implies the final result 
\beqa\label{currentnl}
    \bra a(0)\ket_{\rm LR}
    &=& a(\bra q(0)\ket_{\rm LR})
    + \ell^{-1}\gamma_{\rm s}^{ij}(0^+,0^-)\frc{\p^2}{\p q_m \p q_n}\bra a\ket_{\bra q(0)\ket}C_{mi}(0)C_{nj}(0).
\eeqa
That is, the modification is only at order $\ell^{-1}$, and involves the second derivative wrt densities times the symmetric part of the kernel function around its jump. Note that at order $\ell^{-1}$, we may replace $\bra q(0)\ket$ by $\bra q(0)\ket_{\rm LR}$.

Next we consider the long-range correlation functions, and identify them with the kernel function times $C$ matrices. We evaluate correlation functions (using $q_i^-(x) = q_i(x)-\bra q_i(x)\ket$)
\beqa
    \bra q_k(0)q_l(z)\ket^{\rm c}_{\rm LR}
    &=&\bra q_k(0)q_l(z)\ket^{\rm c}
    +\ell\int \dd x\dd y \,\gamma^{ij}(x,y)
    \bra q_i^-(x)
     q_j^-(y),q_k(0),q_l(z)\ket \n
     &=&\bra q_k(0)q_l(z)\ket^{\rm c}
    +\ell\int \dd x\dd y \,\gamma^{ij}(x,y)
    \Big(
    \bra q_i^-(x)q_j^-(y)q_k(0)q_l(z)\ket^{\rm c}\n &&\qquad
    +\,\bra q_i^-(x)q_k(0)\ket^{\rm c}
    \bra q_j^-(y)q_l(z)\ket^{\rm c}
    +\bra q_i^-(x)q_l(z)\ket^{\rm c}
    \bra q_j^-(y)q_k(0)\ket^{\rm c}\n &&\qquad
    +\,\bra q_i^-(x)q_k(0)q_l(z)\ket^{\rm c}
    \bra q_j^-(y)\ket
    +\bra q_j^-(y)q_k(0)q_l(z)\ket^{\rm c}
    \bra q_i^-(y)\ket
    \Big) \n
    &=&\bra q_k(0)q_l(z)\ket^{\rm c}
    +\ell\int \dd x\dd y \,\gamma^{ij}(x,y)
    \Big(
    \bra q_i(x)q_j(y)q_k(0)q_l(z)\ket^{\rm c}\n &&\qquad
    +\,\bra q_i(x)q_k(0)\ket^{\rm c}
    \bra q_j(y)q_l(z)\ket^{\rm c}
    +\bra q_i(x)q_l(z)\ket^{\rm c}
    \bra q_j(y)q_k(0)\ket^{\rm c}\Big)
    \label{qkqlnl}
\eeqa
where in the second line we used the following cumulant expansion of four-point functions 
\beq
    \bra ab,c,d\ket = \bra a,b,c,d\ket
    + \bra a,c\ket\bra b,d\ket+
    \bra a,d\ket\bra b,c\ket
    +\bra a,c,d\ket\bra b\ket
    +\bra b,c,d\ket\bra a\ket.
\eeq
In \eqref{qkqlnl} the part with $\bra q_i(x)q_j(y)q_k(0)q_l(z)\ket^{\rm c}$ is $O(\ell^{-2})$ so can be neglected in this order. 

Again using that correlations are zero outside a microscopic scale, we have  
\beq
	\bra q_i(x)a(0)\ket^{\rm c} = 0 \ \ (|x|>r_{\rm micro}\ell^{-1}),\quad \int \dd x\,\bra q_i(x)a(0)\ket^{\rm c}
	=-\ell^{-1}\frc{\p}{\p\beta^i}\bra a\ket_{\bra q(0)\ket} + O(\ell^{-2}),
\eeq
so that
\beq
	\ell\int \dd x\dd y\,\gamma^{ij}(x,y)
	\bra q_i(x)q_k(0)\ket\bra q_j(y)q_l(z)\ket^{\rm c}= \ell^{-1}
	\gamma^{ij}(0,z)
    C_{ik}(0) C_{jl}(z)
    +O(\ell^{-2})
\eeq
and finally
\beq
    \bra q_k(0)q_l(z)^{\rm c}_{\rm LR}
    =
    \bra q_k(0)q_l(z)\ket^{\rm c}
    +2\ell^{-1}
	\gamma^{ij}(0,z)
    C_{ik}(0) C_{jl}(z)
    +O(\ell^{-2}).
\eeq
We recovered this way  Eq. \eqref{eq:finalres1} and Eq. \eqref{eq:finalres2}.

Finally, we show that the entropy increase in the NL  state is zero, where the long-range correlation part regularise the entropy growth the local state. Time derivatives gives 
\begin{equation}
    \dot{S} = - \int \dd x \  {\rm Tr \rho_{\rm LR} \log \rho_{\rm LR}} = \int \dd x     \partial \beta^i \langle j_i \rangle - \frac{1}{2} \int \dd x  \dd y   \gamma_s^{ik}(x,y) \frac{ \partial S_{ik}^{\rm reg}}{ \partial t }  
\end{equation}
and using Eq. \eqref{eq:timeevCorr}
\begin{equation}
\dot{S} = \int dx \partial \beta^k (j_k(q_\cdot) + \frac{\gamma_s^{ij}(x,x)}{2}\langle j_k^-,q_i,q_j \rangle ) -  \int \dd x  \gamma_s^{ij}(x,x)    \langle j_k^-,q_i , q_j \rangle \partial \beta^k + \int dx \partial_x   j_s  =0
\end{equation}
where we have denoted with $\partial_x j_s$ the remaining functions which is a total gradient in $x$.

\section{Explicit formulas for the hard rods evolution starting from an initial constant particle density state with local GGE correlations}
The following formulas will derive in detail in \cite{UPCOMING}. We state them here for completeness.

We consider hard rods with size $a$ and an initial state given by a quasi-particle density $\rho(x,p)=\bar\rho h(x,p)$ and GGE-correlations:
\beqa
    L\langle \rho(x,p)\rho(y,q)\rangle \upd{c} = \delta(x-y)[\rho(x,p)\delta(p-q)+\rho(x,p)\rho(y,q)(-a+a^2\bar\rho)],
\eeqa
where $\int\dd{p}h(x,p)=1$ means we consider states of constant particle densities (the case of non-constant particle density can be treated as well, but gives rise to more complicated formulas). The quasi-particle density $\rho(t,x,p) = \rho\ind{E}(t,x,p) + \tfrac{1}{L}\rho\ind{D}(t,x,p) + \mathcal{O}(1/L^2)$ after time $t$ is given by:
\begin{align}
    \rho\ind{E}(t,x,p) &= \frac{\rho(0,X^{-1}(t,x,p),p)}{\frac{\mathrm{d}X(t,x,p)}{\mathrm{d}x}}\\
    \rho\ind{D}(t,x,p) &= -\partial_x (\rho\ind{E}(t,x,p)\Delta X(t,X^{-1}(t,x,p),p)) + \tfrac{1}{2}\partial_x^2(\rho\ind{E}(t,x,p)V(t,X^{-1}(t,x,p),p)),
\end{align}
where
\begin{align}
    X(t,x,p) &= x+pt+a\int\dd{y}\dd{q}\rho(y,q)(\theta(x-y+\tfrac{p-q}{1\upd{dr}})-\theta(x-y))
\end{align}
is the GHD characteristic of a particle starting at $x,p$ ($1\upd{dr} = 1-a\bar\rho$),
\begin{align}
    \Delta X(t,x,p) &= a^2 {1\upd{dr}}\bar{\rho}\int\dd{p_2}h(x+\tfrac{p-p_2}{1\upd{dr}}t,p_2)\sgn(p-p_2) + \tfrac{a^3}{2}\bar\rho^2\int\dd{p_2}\partial_x h(x+\tfrac{p-p_2}{1\upd{dr}}t,p_2)\tfrac{|p-p_2|}{1\upd{dr}}t\\
    &+ \tfrac{a^3}{2}\bar\rho^2\int\dd{p_2}h(x+\tfrac{p-p_2}{1\upd{dr}}t,p_2)\sgn(p-p_2) +\tfrac{a}{2}\bar\rho\int\dd{p_2}h(x,p_2)(-2a+a^2\bar\rho) \sgn(p-p_2)
\end{align}
is the $1/L$ correction to the expected position of the particle after time $t$ and
\begin{align}
    V(t,x,p) &= 2a^3\bar\rho^31\upd{dr}\int\dd{x_2}\dd{p_2}\dd{p_3}h(x_2,p_2)h(x_3,p_3)(\theta(x-x_2+\tfrac{p-p_2}{1\upd{dr}}t)-\theta(x-x_2)) {\bf 1}_{(x,x_3)}(x_2)\Big|_{x_3=x+\tfrac{p-p_3}{1\upd{dr}}t}\\
	&+a^4\bar\rho^3\int\dd{p_2}\dd{p_3}h(x,p_2)h(x,p_3)\theta((p-p_2)(p-p_3))\tfrac{|p-p_2|\wedge |p-p_3|}{1\upd{dr}}t\\
	&+a^2(-2a+a^2\bar\rho)\bar\rho^2\int\dd{x_2}\Big[\int\dd{p_2} h(x,p_2)(\theta(x-x_2+\tfrac{p-p_2}{1\upd{dr}}t)-\theta(x-x_2))\Big]^2\\
    &+ a^2\bar\rho\int\dd{x_2}\dd{p_2}h(x_2,p_2) \Big[\theta(x-x_2+\tfrac{p-p_2}{1\upd{dr}}t)-\theta(x-x_2)\Big]^2
\end{align}
is the variance of the trajectory. Here ${\bf 1}_{(x_1,x_3)}(x_2)= \theta(x_3-x_2)-\theta(x_1-x_2)$ and $a\wedge b = \mathrm{min}(a,b)$.
Furthermore the correlations at time $t$ are given by:
\begin{align}
    \langle \rho(t,x,p)&\rho(t,y,q)\rangle\upd{c} = \partial_x\partial_y \Big[
    \int\dd{p'}\dd{q'}(\delta(p-p')-a\rho\ind{E}(t,x,p))(\delta(q-q')-a\rho\ind{E}(t,y,q))\\
    &\times 
    \int\dd{p''}\dd{q''}(\delta(p'-p'')+a\tfrac{\rho\ind{E}(x',p')}{1\upd{dr}})(\delta(q'-q'')+\tfrac{\rho\ind{E}(y',q')}{1\upd{dr}})\\
    &\times \Big(\int_{-\infty}^{x'\wedge y'}\dd{z}\rho(z,p'')\delta(p''-q'') + \rho\ind{E}(z,p'')\rho\ind{E}(z,q'')(-2a+a^2\bar{\rho}) \Big)\Big|_{x'=\hat{X}^{-1}(0,\hat{X}(t,x)-p't),y'=\hat{X}^{-1}(0,\hat{X}(t,y)-q't)}\Big],
\end{align}
with $\hat{X}(t,x) = x-a\int_{-\infty}^x\dd{y}\dd{p}\rho\ind{E}(t,y,p)$ and $\hat{X}^{-1}(t,\hat{x})$ is its inverse in $x$. These formulas are evaluated numerically.

\section{Cumulant expansion}
Here we review the cumulant expansion of an observable in the large deviation setting. For simplicity we take a single random variable; the extension to many random variables is immediate. Its cumulant $\kappa_{n}(X)$ of order $n\geq 1$ may by defined by setting $\kappa_1(X) = \bra X\ket$ and
\beq
    \bra e^{(X - \bra X\ket) t}\ket
    = e^{F_{\geq 2}( t)},\quad F_{\geq 2}( t) = \sum_{n=2}^\infty 
    \kappa_n(X)\frc{t^n}{n!}.
\eeq
Taylor expanding a smooth function around $a$ as $f( x) = e^{(x -  a) \p}f|_{ a}$, we have
\beq
    \bra f(X)\ket = 
    \bra e^{( X - \bra X \ket) \p}\ket f|_{\bra X\ket}
    = e^{F_{\geq 2}(\p)}f|_{\bra X\ket}.
\eeq
Expanding the exponential, this gives is an expansion of the average of $f(X)$ in terms of cumulants of $X$ (organised in cumulants' orders). In the large deviation setting, one has $\kappa_n(X) \sim 1/\ell^{n-1}$ for some large scale $\ell$. Then the above gives
\beq
    \bra f(X)\ket = f(\bra X\ket) + \frc{f''(\bra X\ket)}2 \kappa_2(X) + O(\ell^{-2}).
\eeq
Alternatively, one may also define a cumulant of order $n$ via the moment-to-cumulant relation (adapting the notation in a natural way),
\begin{align}
    \langle X_1\cdots X_n\rangle &= \sum_\pi \prod_{B\in \pi} \kappa(X_i:i \in B),
\end{align}
where $\pi$ runs over all partitions of $\{1,\ldots,n\}$ and $B$ over all blocks in the partition $\pi$. In the large deviation setting, from this one can show that, for centered random variables $Y_1,\ldots,Y_n$ with $\langle Y_i\rangle=\kappa(Y_i)=0$, we then have
\begin{align}
    \langle Y_1\cdots Y_n\rangle &\sim 1/\ell^{\lceil\tfrac{n}{2}\rceil}.
\end{align}
Define $Y=X-\langle X\rangle$, Taylor expanding we obtain
\begin{align}
    \langle f(X)\rangle &= \langle f(\langle X\rangle+Y)\rangle = \sum_{n=0}^\infty \frac{f^{(n)}(\langle X\rangle)}{n!} \langle Y^n\rangle.
\end{align}
Using the fact that cumulants of order 2 and above are invariant under constant shifts of the random variable, the moment-to-cumulant relation again gives the explicit expansion in cumulants.
\end{document}